# Zero-Group-Velocity acoustic waveguides for high-frequency resonators

C. Caliendo[*,1], M. Hamidullah[1]

[1]Institute of Photonics and Nanotechnologies, IFN-CNR, Via Cineto Romano 42, 00156 Rome, Italy

*Corresponding author; email: cinzia.caliendo@cnr.it

**Abstract.** The propagation of the Lamb-like modes along a silicon-on-insulator (SOI)/AlN thin supported structure was theoretically studied in order to exploit the intrinsic zero group velocity (ZGV) features to design electroacoustic resonators that don't require metal strip gratings or suspended edges to confine the acoustic energy. The ZGV resonant conditions in the SOI/AlN composite plate, i.e. the frequencies where the mode group velocity vanishes while the phase velocity remains finite, were investigated in the frequency range from few hundreds of MHz up to 1900 MHz. Some ZGV points were found that show up mostly in low-order modes. The thermal behaviour of these points was studied in the -30 to 220 °C temperature range and the temperature coefficients of the ZGV resonant frequencies (TCF) were estimated. The behaviour of the ZGV resonators operating as gas sensors was studied under the hypothesis that the surface of the device is covered with a thin polyisobutylene (PIB) film able to selectively adsorb dichloromethane ($CH_2Cl_2$), trichloromethane ($CHCl_3$), carbontetrachloride ($CCl_4$), tetrachloroethylene ($C_2Cl_4$), and trichloroethylene ($C_2HCl_3$). at atmospheric pressure and room temperature. The sensor sensitivity to gas concentration in air was theoretically estimated for the first four ZGV points of the inhomogeneous plate. The feasibility of high-frequency, low TCF electroacoustic micro-resonator based on SOI and piezoelectric thin film technology was demonstrated theoretically.

Keywords: zero group velocity; Lamb-like modes; SOI; AlN; acoustic waveguides.

## 1. Introduction

Lamb waves are acoustic guided modes that propagate in finite thickness plates and are strongly dispersive [1]. For some branches of the dispersion curves, a strong resonance can be found that occurs at the frequency minimum: at this frequency, a zero-group velocity (ZGV) Lamb mode occurs that is characterized by a vanishing group velocity combined with a non-zero wave number [2]. More specifically, the zero group velocity is due to the interference of two modes with the same frequency and mode shape, and propagating with equal phase velocity in opposite directions. As a result, a stationary non-propagating mode is obtained that corresponds to a local resonance in the response spectrum of the plate where the Lamb modes travel. ZGV modes have been proposed for a wide range of applications such as the estimation of the Poisson's ratio [3, 4], the measurement of the thickness of plates [5, 6], the probing of interfacial stiffness between two plates [7 -10], to cite just a few. The most of the available literature refers to ZGV modes along tungsten [4], aluminium [6], glass [9], duralumin

[5, 10], poly methyl methacrylate [11] and metal plates bonded with submillimetric glue layer [9]. Although the literature concerning ZGV Lamb modes is rather extensive, to the authors' knowledge, only a few studies consider the topic of ZGV electro-acoustic resonators on thin suspended acoustic waveguides: in the references [12 ,13] the first symmetric Lamb $S_1$ mode ZGV resonator based on a c-AlN thin suspended membrane (2.5 μm thick and wavelengths of about 8.0 - 8.8 μm) is theoretically studied and experimentally verified. The impact of the free edges shapes (flat or biconvex) of the piezoelectric suspended rectangular membrane, the tether-to-plate angle (90º or 59° for "butterfly" shaped plate), electrode configurations, materials, and thicknesses on the performances of the AlN-based Lamb wave resonators have been investigated in references [14 - 17]. As opposed to the ZGV resonators, the resonators based on a piezoelectric thin suspended membrane with free edges require a more complicated micromachining technology to acoustically isolate the resonator and mechanically couple it to the substrate by two anchors. Indeed, the ZGV points are associated with an intrinsic energy localization: the energy confinement is a natural consequence of the selected acoustic mode thus reducing the technological complexity with respect to that required by the free edges resonators. The characteristics of the ZGV resonators, such as their phase velocity, resonant frequency and Q factor, are affected by the properties of the layer materials, the thickness of the plate, as well as by the electrical boundary conditions. Silicon on insulator (SOI) technology refers to the use of a layered silicon–insulator–silicon substrate in place of conventional bulk silicon substrates in microelectronics semiconductor manufacturing. SOI substrates are compatible with a broad spectrum of technologies that are used in conventional areas of micro-electronics: they provide the potential for high-performance electronics with multiple integrated functions, such as actuators and sensors, and are suited for the production of large-scale integrated systems [18, 19]. AlN is a piezoelectric material with unique properties such as a wide band gap, high thermal conductivity, a low thermal expansion coefficient, high breakdown dielectric strength, and the highest acoustic wave velocity among the piezoelectric materials that can be grown in thin film form [20]. SOI/AlN-based suspended membranes offer process robustness and design flexibility to provide devices with multiple ZGV frequencies implemented on single-chip platforms. This paper provides a theoretical demonstration of monolithically integrated piezoelectric MEMS RF resonators using SOI substrates and the piezoelectric films technology, suitable for frequency control and sensing applications.

## 2. LAMB-like waves
Lamb modes are acoustic waves that propagate in finite thickness homogeneous isotropic plates. They are distinguished as symmetric and antisymmetric modes, $S_n$ and $A_n$, where n is the mode order. These modes have two particle displacement components, $U_1$ and $U_3$: $U_1$ is parallel to the wave propagation direction $x_1$, and $U_3$ is normal to the plate surface, hence parallel to $x_3$. $U_1$ ($U_3$) is symmetric while $U_3$ ($U_1$) is antisymmetric respect to the mid-plane of the plate for the symmetric (antisymmetric) modes. If the finite thickness plate is non homogeneous, i.e. multi-layered or composite plate, the distinction between symmetric and antisymmetric modes is no more valid since the field profiles of the propagating modes have no longer a symmetry around the mid-plane of the plate. This is the case under study where the waves propagate along a multi-layered composite plate including a piezoelectric AlN layer on top of a SOI suspended membrane [21].
The fundamental symmetric and antisymmetric modes can be considered as quasi-$S_0$ and quasi-$A_0$, (q$S_0$ and q$A_0$) just for a limited plate thickness range, while all the other modes will be generally labelled as Lamb-like modes as the distinction between mode types is somewhat artificial. The higher order modes will be generically distinguished by a number in the order in which they appear along the frequency axis.

*2.1. Phase and group velocity dispersion curves*
The propagation of Lamb waves along a SOI/AlN thin suspended membrane is here investigated. The membrane consists of a piezoelectric AlN layer, 1 μm thick, on top of a SOI suspended membrane with a 10 μm Si layer and a 1 μm thick $SiO_2$ box layer. Figure 1 shows the schematic of the resonator including the array of electrodes on top of the piezoelectric layer. This device can be obtained by standard technological processes, such as the backside SOI/AlN micro-machining process for the fabrication of suspended membranes. In this case the silicon dioxide box plays the role of a back-etching stop layer, allowing the release of a Si/SiO$_2$/AlN suspended membrane.

The total composite plate thickness is H = h$_{SiO2}$ +h$_{Si}$ +h$_{AlN}$, where h$_{SiO2}$ is the SiO$_2$ layer thickness, h$_{Si}$ is the Si layer thickness, and h$_{AlN}$ is the piezoelectric AlN layer thickness. The Lamb waves propagation can be excited and detected by use of interdigitated transducers (IDTs), as for the surface acoustic waves (SAWs). The wavelength of the acoustic wave, λ, is set by the pitch of the interdigital transducer (IDT), while the number of interdigitated electrodes of both the input and output IDTs is equal to $N = \sqrt{\pi/4K^2}$, as required to obtain the minimum insertion loss and the maximum frequency bandwidth of the resonator implemented on the SOI/AlN plate; K$^2$ is the electroacoustic coupling coefficient that depends on the layers thickness and electrical boundary conditions. In the present simulations, the IDTs metallization ratio is supposed to be equal to 1.

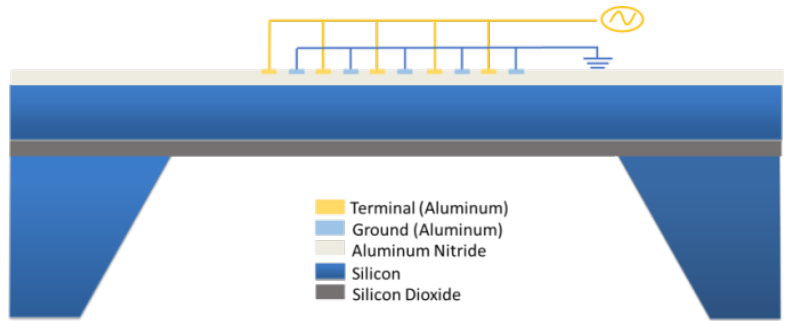

**Figure 1.** Schematic of the Lamb mode resonator on AlN/Si/SiO$_2$ suspended membrane with a thick Si rigid frame.

The phase velocity dispersion curves of the Lamb-like modes travelling along the SOI/AlN composite plate were calculated and plotted in figure 2a vs H/λ, being H the total waveguide thickness. The labels near each curve show the corresponding modes order: the fundamental modes are labelled as qS$_0$ and qA$_0$, while the higher order modes are indicated with a progressive number. The group velocity dispersion curves of the composite plate were calculated according to the formula $v_{gr}^n = v_{ph}^n \left(1 + \frac{H}{\lambda} \frac{\partial v_{ph}^n}{\partial H/\lambda} \frac{1}{v_{ph}^n}\right)$, where n is the mode order, and are shown in figure 2b. The curves of figure 2b are distinguished on the basis of the color code adopted in figure 2a.

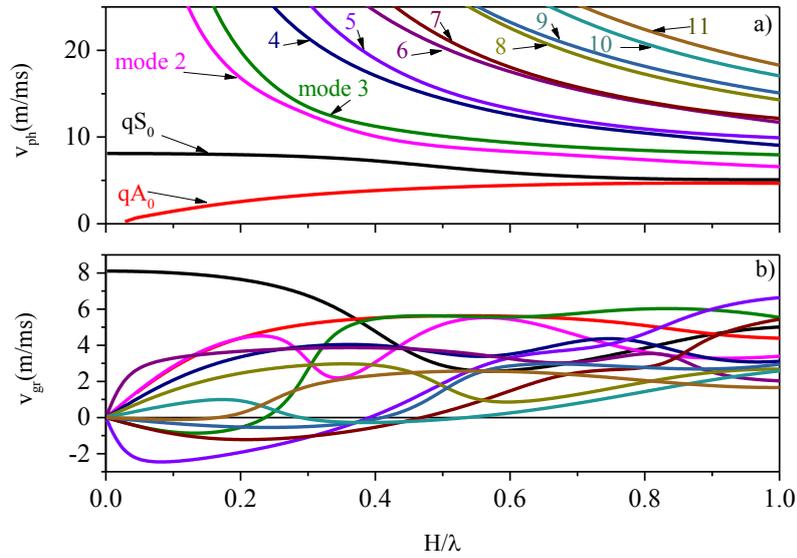

**Figure 2.** a) The phase velocity and b) the group velocity vs H/λ of the Lamb modes in the composite plate. Same colored curves in both figures belong to the same mode.

As it can be seen in figure 2a, the phase velocity of the lowest two modes, $qS_0$ and $qA_0$, (the black and red curves) are continuous with frequency, while the higher order modes originate at a cut-off frequency at which the phase velocity is infinite and the plate vibrates in longitudinal or shear thickness mode resonance. For the $qS_0$ mode, the group velocity (the black curve of figure 2b) is equal to the phase velocity in the low-frequency limit but there after falls below the latter, reaching a minimum near the H/λ equal to 0.57. The high-frequency asymptotic limit is the Rayleigh wave velocity. The group velocity of the $qA_0$ mode (the red curve in figure 2b) rises rapidly from its low-frequency limit of zero and, at H/λ of about 0.37 and 0.9, it equals that of the $S_0$-like mode The group velocity of the higher order modes vanishes at $k = 0$, giving rise to a thickness resonance at the cut-off frequency. At these cut off frequencies, multiple reflections between the top and bottom surfaces of the plate result in a *thickness resonance*. Some branches, corresponding to the ''backward-wave'' propagation, occur in the negative-slope region where group velocity and phase velocity have opposite signs. For negative group velocities, the direction of propagation of wave energy and that of wave phase are opposite. In addition to the thickness mode resonances, some Lamb wave resonances, referred to as the zero group velocity (ZGV) resonances, occur at the frequency values at which some high order modes exhibit null group velocity and finite phase velocity. Figure 3 shows a magnification of figure 2b: one can easily observe that all modes have one ZGV point, except the 10 mode that has two ZGV points.

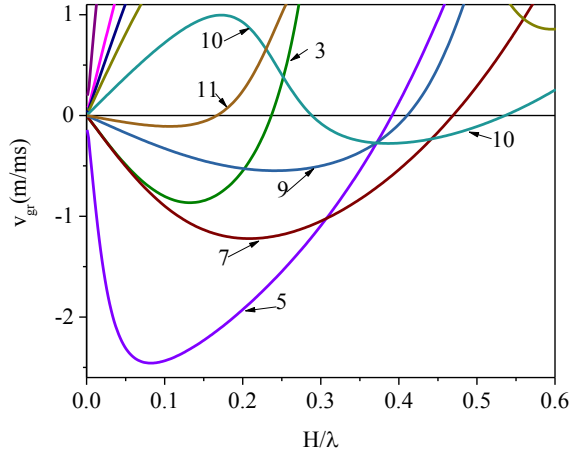

Figure 3: A magnification of figure 2b that shows the single and double ZGV points.

The frequencies of the ZGV points are listed in Table 1 together with the corresponding wavelength values. As an example, the $S_1$-like mode ($qS_1$) phase and group velocity dispersion curves in the vicinity of the ZGV resonance frequency are shown in Figure 4 for a composite plate of thickness H = 12 μm. It can be seen that the group velocity is zero at two frequencies corresponding to the thickness resonance frequency (with infinite value of $v_{ph}$, hence $k = 2\pi/\lambda = 2\pi f/v_{ph} = 0$) and the ZGV resonance frequency. The origin of the latter resonance is due to the behaviour of the dispersion curve at frequencies close to the ZGV resonance, where two branches with different slopes can be distinguished: the upper branch with positive slope, corresponding to the negative $v_{gr}$ values, and the lower $v_{ph}$ branch with negative slope, corresponding to the positive $v_{gr}$ values [22]. The upper branch is backward propagating because the phase and group velocities of the mode are of opposite sign, and thus the direction of propagation of the acoustic energy is opposite to the wave vector. On the contrary, the phase and group velocities of the lower branch have the same sign and this branch is considered to be a true mode. In the vicinity of $qS_1$ ZGV resonance frequency, the absolute value of the group velocity of the two branches approach zero, while their phase velocities have the same non null value. These modes interfere and form the standing $qS_1$ ZGV resonance.

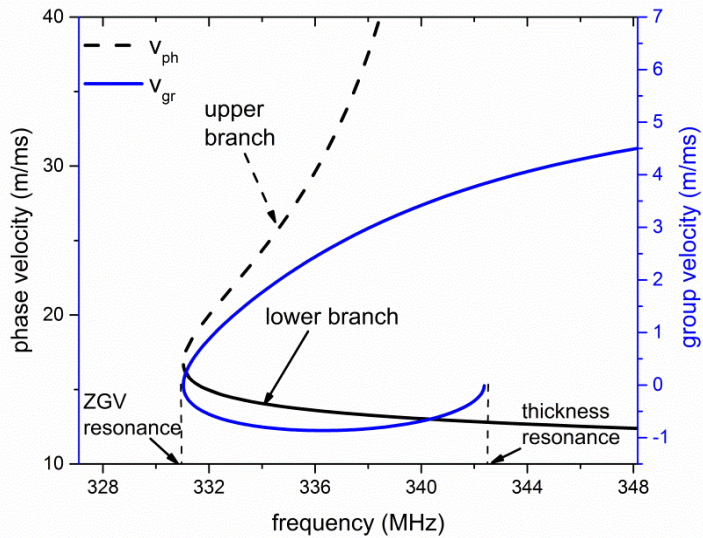

**Figure 4.** Phase velocity and group velocity dispersion curves for the $qS_1$ mode in the vicinity of the $qS_1$ ZGV3 resonance.

As it can be noticed in figure 3, there is a mode whose dispersion curve has two turning points at which the group velocity vanishes, corresponding to H/λ = 0.54 and 0.29 in the frequency spectrum range up to 1900 MHz. The dispersion curve of this mode undergoes a change of sign in its slopes twice, thus leading to two ZGV points, as shown in figure 5 where the $v_{ph}$ dispersion branches with positive and negative slope have been coloured in red and black, respectively. The phase and group velocities are oppositely directed in the frequency range between these two ZGV points. This phenomenon is not rare as described in reference [23] where some examples of multiple ZGV points are given.

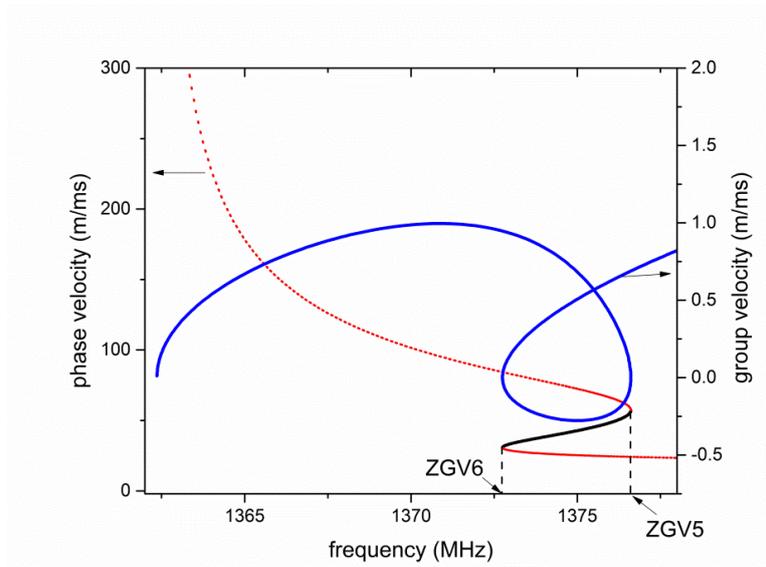

**Figure 5.** Phase velocity and group velocity dispersion curves in the vicinity of the ZGV10a and ZGV10b resonances.

Figures 6a-c show, as an example, a comparison between the $v_{gr}$ vs frequency curves of three different plates: a) an AlN plate, b) a Si plate, c) and the SOI/AlN composite plate.

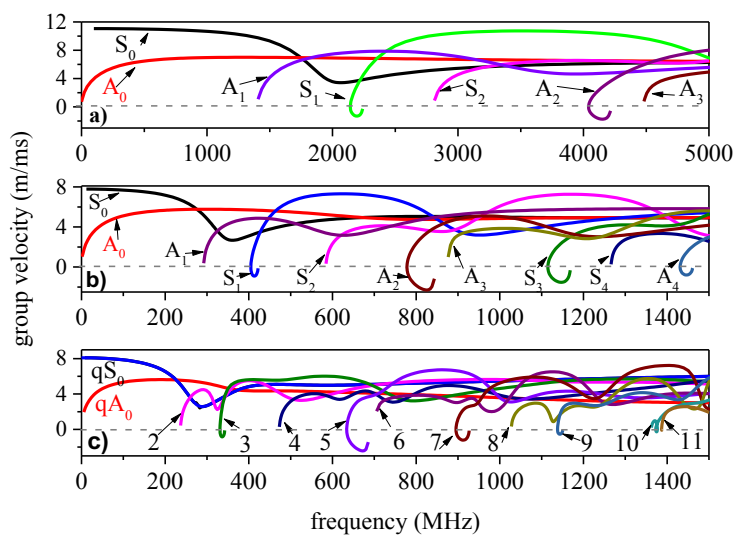

**Figure 6.** The group velocity vs frequency curves of the modes propagating along the a) c-AlN plate 2.5 μm thick; b) Si plate 10 μm thick; c) SiO$_2$/Si/AlN plate, 1μm/10μm/1μm thick.

The distinction between symmetric and antisymmetric Lamb modes in figures 6a and 6b, is still possible for one-material plate, while this distinction is somewhat artificial for a composite plate. Thus the dispersion curves of figure 5c have been generically distinguished by a number in the order in which the modes appear along the frequency axis after the two fundamental modes (the 0 and 1 mode), labelled qA$_0$ and qS$_0$. From figure 6 it can be noticed that the Lamb mode ZGV phenomenon is strictly related to the plate composition, i.e. the layers material and thickness. Figure 6a refers to an AlN plate, 2.5 μm thick, that exhibits only two ZGV points, the first order symmetric Lamb mode (named ZGVS$_1$) and the second order anti-symmetric mode (named ZGVA$_2$). ; Figure 6b refers to a Si plate, 10 μm thick, that exhibits several ZGV points. Figure 6c refers to a three layer materials composite plate, SiO$_2$/Si/AlN, with thicknesses equal to 1, 10 and 1 μm, respectively: as it can be seen, the frequencies of the ZGV points are lower than those of the bare Si plate due to the presence of the SiO$_2$ and AlN layers, while the numbers of these ZGV points remains the same. The phase and group velocity dispersion curves were calculated by using the software DISPERSE [24]; the AlN, Si and SiO$_2$ material constants are from 25, 26.

## 3. ZGV acoustic waveguides for high-frequency microwave resonators

*3.1. The coupling efficiency*

One or two-ports SAW resonators (SAWRs) employ one or two IDTs to generate and receive the SAW, and two or three Bragg acoustic mirrors which reflect the SAW and generate a standing wave between the reflectors. The SAW is reflected and confined between the two IDTs. Several arrangements and geometries of the reflectors are studied to balance performances issues, cost, and manufacturability in reference 25. Thin-film plate acoustic resonators can be developed in two different topologies with respect to the acoustic wave reflection: one is based on the reflection from a periodic grating [28], like SAW resonators, thus requiring the study of the device aperture, grating reflectivity, and synchronization to optimize the device response. The other is based on the reflection from the suspended free edges of the thin plate: the Lamb modes are reflected by the suspended free edges at both sides of the piezoelectric thin plate, and consequently the standing waves form inside the thin plate. The loss due to mode conversion upon reflection at the suspended free edges can be significantly reduced by propagating the lowest-order Lamb wave modes [29]. In reference 14 it is demonstrated that the suspended biconvex edges of the AlN membrane can more efficiently confine the mechanical energy in the S$_0$ Lamb mode resonator and reduce the energy dissipation at the edges through the support tethers than the flat edges resonator. The fabrication process flow for manufacturing the free edges suspended membrane is quite complicate, as described in reference 29. The ZGV points of a Lamb mode device are associated with an intrinsic energy localization: this fact enables the design of acoustic micro-resonators employing only one IDT and no reflectors, thus reducing both the device size and the technological complexity, while the energy confinement is a natural consequence of the selected acoustic mode. In the SOI/AlN plate, four piezoelectric coupling configurations can be obtained by placing the IDT at the SOI/AlN interface (SOI-Transducer-Film, STF) or at the AlN surface (SOI-Film-Transducer, SFT), further including a floating metal electrode onto the AlN layer side opposite to that where the IDT is located (SOI-Transducer-Film-Metal and SOI-Metal-Film-Transducer, STFM and SMFT). The four configurations are depicted in figure 7.

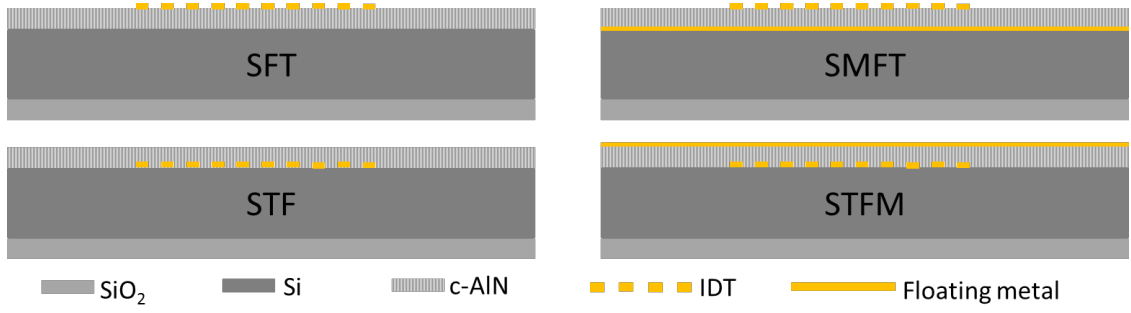

**Figure 7.** The four coupling configurations of the ZGV resonator.

The electroacoustic coupling coefficient, $K^2$, physically represents the IDT electrical to acoustic energy conversion efficiency: it is frequency dispersive and strongly affected by the electrical boundary conditions. The $K^2$ can be enhanced by placing a metal floating electrode onto the AlN side opposite to that where the IDT is located. FEM analysis was performed using COMSOL 5.2 to calculate the $K^2$ as $K^2 \approx 2 \cdot \frac{(v_f - v_m)}{v_f}$. where $v_f$ and $v_m$ are the velocities along the electrically open and shorted surfaces of the AlN film. In table 1 the resonant frequency, the wavelength λ, the phase velocity, and the $K^2$ of the four coupling configurations are summarized for each ZGV point of the $SiO_2$-Si-AlN composite plate (with thicknesses equal to 1 μm, 10 μm and 1 μm, respectively).

**Table 1.** the ZGV mode order, the wavelength, the modes frequency, and the $K^2$ of the four coupling configurations.

| ZGV mode order | λ (μm) | Freq. (MHz) | SFT (%) | STF (%) | SMFT (%) | STFM (%) |
|---|---|---|---|---|---|---|
| 3 | 50 | 328.29 | 0.001 | 0.0035 | 0.049 | 0.051 |
| 5 | 30 | 626.53 | 0.013 | 0.027 | 0.066 | 0.080 |
| 7 | 25.5 | 885.86 | 0.035 | 0.058 | 0.051 | 0.075 |
| 9 | 29 | 1133.10 | 0.058 | 0.080 | 0.036 | 0.058 |
| 10a | 41.5 | 1371.90 | 0.077 | 0.073 | 0.005 | 0.001 |
| 10b | 22 | 1367.30 | 0.074 | 0.093 | 0.009 | 0.028 |

The 3 and 5 ZGV mode order of table 1 refer to the $qS_1$ and $qA_2$ modes, respectively. The calculations were performed assuming the IDTs and floating electrode as infinitely thin and perfectly conductive. It is worth noting that the $K_2$ of the ZGVRs is affected by the IDTs and floating electrode thicknesses, the IDTs metallization ratio, the combination of electrode materials for both the IDTs and floating electrode, and the electrical boundary conditions (i.e., floating or grounded metal electrode opposite the IDTs). A proper design of the coupling configuration could result in a substantial improvement of the $K_2$ value respect to the values listed in table 1 [13]. At the present, this type of investigation is out of the scope of the paper but will be matter of study of our future work.

*3.2. ZGV energy confinement*

The ZGVR consists of two area, as shown in figure 1: one metalized (under the IDT or in between the IDT and the floating electrode) and one bare (the outside region), which exhibit slightly different dispersion characteristics. To avoid the coupling between the modes excited in the active device region to those from the area surrounding the resonator, the frequency of the ZGV points in the outside region must be higher than that in the active region. This is readily accomplished by the IDT and floating electrode mass loading effect that lowers the wave velocity in the active region with respect to that in the outside bare region, as verified modelling the resonator as an aluminium (Al film 0.2 μm thick) continuously metalized AlN layer positioned on top of the SOI substrate. In Figures 8a-d, the frequency vs H'/λ curves of the first four modes that exhibit a ZGV point are shown, being $H' = h_{SiO2} + h_{Si} + h_{AlN} + h_{Al}$. The modes are the following: qS$_1$ (mode 3), qA$_2$ (mode 5), mode 7, and mode 9, also indicated as ZGV3, ZGV5, ZGV7 and ZGV9. Four curves are shown for each mode that refer to different layer electrical boundary conditions of the AlN layer. When both the AlN sides are free or metallized the possible configurations are the free-free (SiO$_2$/Si/AlN) and met-met (SiO$_2$/Si/Al/AlN/Al); when one AlN side is free and the other is metallized, the possible configurations are the free-met (SiO$_2$/Si/AlN/Al) and met-free (SiO$_2$/Si/Al/AlN). The calculated data are affected only by the mass-loading effect from the Al layer, while the electric loading effect was ignored. The four curves of figures 8a-d still exhibit zero slope points and thus the principle of ZGV energy trapping is satisfied; moreover, the inability of these modes in the active area to transfer part of their energy to the outside region via propagating modes, confirms the energy self-confining nature of the proposed ZGV Lamb wave coupling configurations.

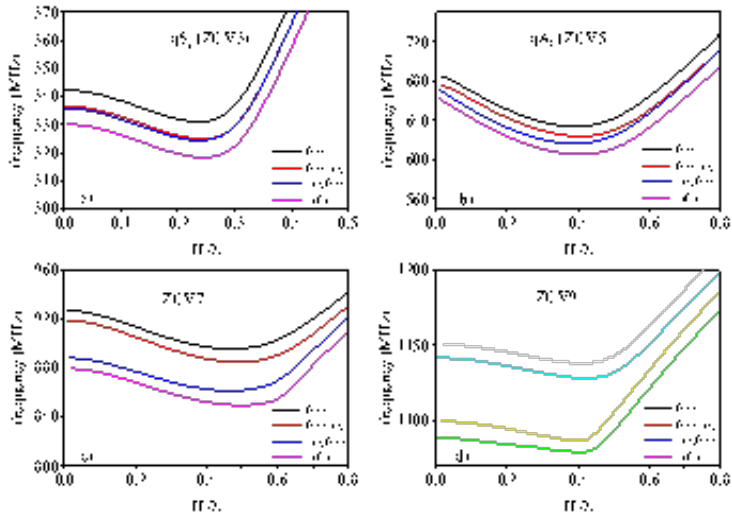

**Figure 8.** The frequency dispersion curves, $f$ vs $H'/\lambda$, of the a) ZGV3, b) ZGV5, c) ZGV7 and d) ZGV9 modes.

FEM analysis was performed using COMSOL Multiphysics to explore the field shape of the ZGV points in the composite waveguide. Initially, the plate-guided modes were identified by frequency-response analysis of a 2D composite plate with the IDT of the ZGV resonator (ZGVR) assumed to have 5 finger pairs, as shown in figure 1. For simplicity, the presence of the silicon frame was not accounted in the simulations and the total length of composite plate is 20·λ. Traction free boundary conditions were selected for the top and bottom sides of the composite plate, while continuity boundary condition were selected for the right and left end sides of the waveguide. The geometrical parameters of the ZGVR structure used in the simulation are given in Table 2.

**Table 2.** The geometrical parameters of the ZGVR structure used in the COMSOL simulation.

| IDT Al electrode thickness | 100 nm |
|---|---|
| AlN layer thickness | 1 μm |
| Si layer thickness | 10 μm |
| SiO$_2$ layer thickness | 1 μm |
| Acoustic wavelength, λ | 50 μm, 30 μm, 25.5 μm, 29 μm, 41.5 μm, 22 μm |

As an example, figure 9a shows the field profile of a *propagating* mode, the qS$_0$ mode, while figures 9b and c show two *non-propagating* modes, the ZGV3 and ZGV5 modes. The total displacements of the modes shown in figures 9a-c were determined by an eigenfrequency 2D FEM analysis with applied boundary conditions: the colour density is representative of the relative particle displacement. The colour bar

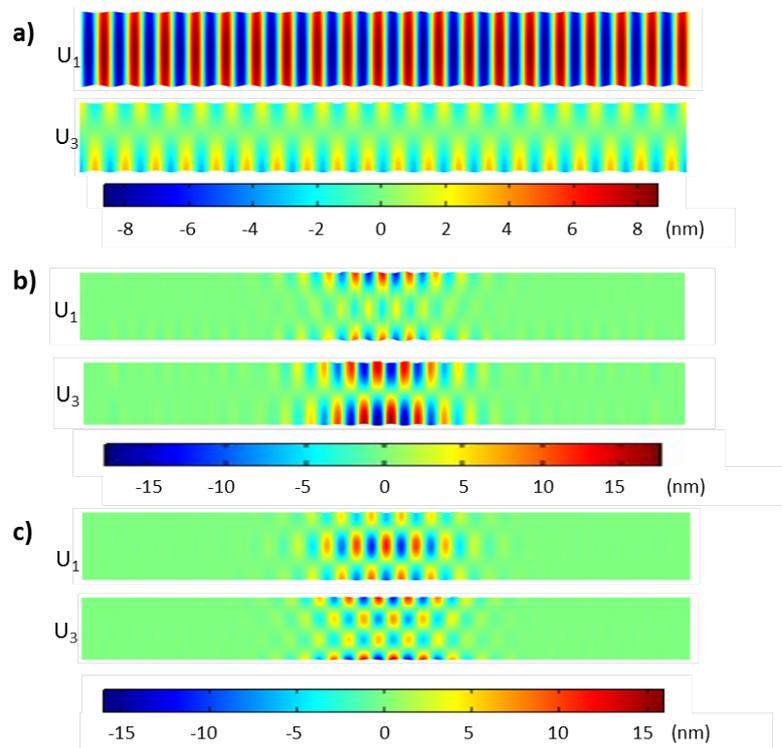

**Figure 9.** The field profile of a) a propagating mode, the qS$_0$ mode, at frequency 155 MHz and λ = 50 μm; b) non-propagating ZGV3 mode for λ = 50 μm, c) non-propagating ZGV5 mode for λ = 30 μm.

As it can be seen in figures 9a-c, the acoustic field shape of the qS$_0$ mode is uniformly distributed under the metal electrodes, as well as in the outside bare region, while the acoustic field of the ZGV3 and ZGV5 modes are localized in the active region of dimension approximately equal to 5 wavelengths. The acoustic field of the higher order ZGV modes are shown in figures 10a-d: the modes are here labelled by using a progressive number since they field shape can be no more identified as quasi-symmetric and quasi-antisymmetric modes.

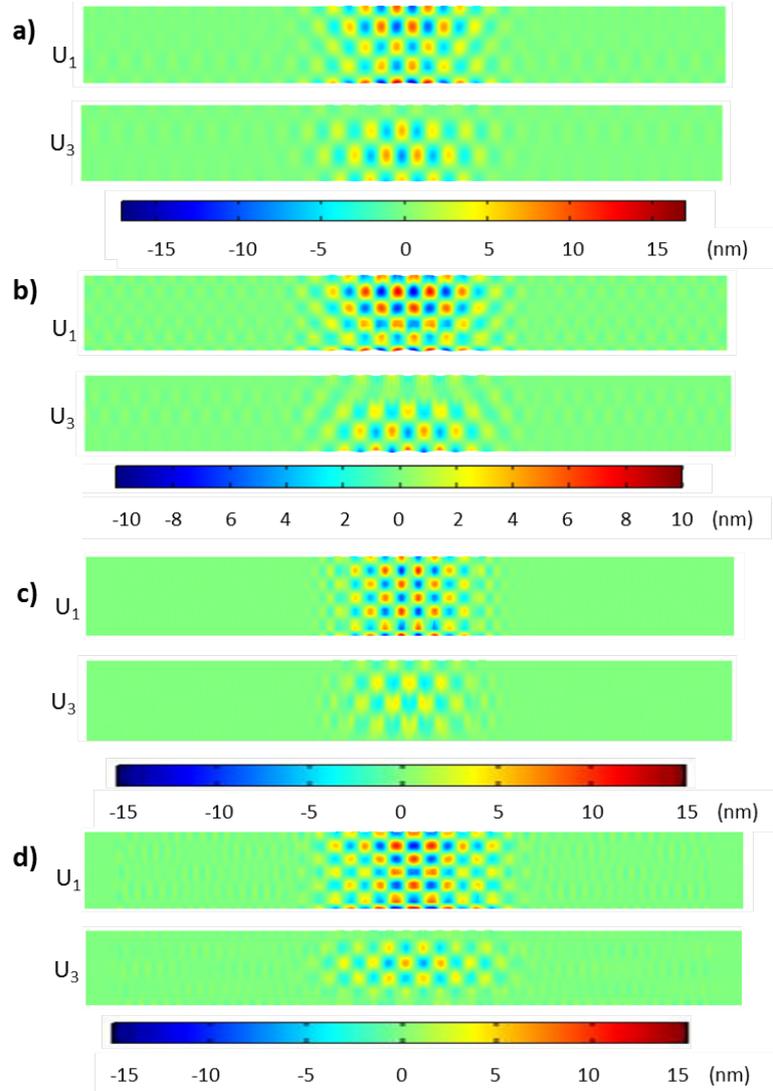

**Figure 10.** The field profile of higher order non-propagating modes, a) ZGV7 mode for λ=25.5μm, b) ZGV9 mode for λ=29μm, c) ZGV10a mode for λ=41.5μm, and d) ZGV10b mode for λ=22μm.

## 4. Thermal behaviour of the ZGV points

The temperature coefficient of the frequency (TCF) represents the relative frequency shift per unit temperature change. Numerical analysis was performed using COMSOL software to evaluate the ZGV resonant frequency $f$, at different temperatures T, from -30 to 220°C. The values of the elastic constants $c_{ij}$ of the three materials (Si, $SiO_2$ and AlN) were estimated for different temperature values according to the formula $c_{ij}(T) = c_{ij}^{20°C}[1 + Tcc_{ij}(T - 20°C)]$, being $Tcc_{ij}$ the temperature coefficient of $c_{ij}$; the temperature dependence of the mass density ρ was calculated by $\rho(T) = \rho(20°C) \cdot [1 - (\alpha_{11} + \alpha_{22} + \alpha_{33})(T - 20°C)]$, being $\alpha_{ii}$ the thermal expansion coefficient (TCE) along the *ii* direction, *ρ(20°C)* the mass density at 20°C. Table 3 lists the mass density, the thermal expansion coefficient, the elastic, piezoelectric, dielectric constants, and the temperature coefficients used in the present calculations [25, 26]. All the theoretical calculations were performed in the lossless materials approximation. No temperature dependent data for the Al IDTs was included because the electrodes are supposed to be much thinner than the smallest layer. The simulation method applied in Comsol included the following steps: definition of the basic structure; definition of the boundary conditions; integration of the temperature dependent material constants; definition of the plate thermal expansion coefficient; parametric investigation of the structure to evaluate the TCF.

**Table 3**: The mass density ρ, the thermal expansion coefficients $\alpha_{ii}$, the elastic constants $c_{ij}$, the piezoelectric constants $e_{ij}$, the dielectric constants $\varepsilon_{ij}$, and the temperature coefficients of the elastic constants $Tcc_{ij}$ of Si, $SiO_2$ and AlN; the data were extracted from references 24 and 25.

| AlN | |
|---|---|
| $c_{11}$ ($10^{11}$ dyn/cm$^2$); $Tcc_{11}$ (ppm/°C) | 3.45; -80 |
| $c_{12}$ ($10^{11}$ dyn/cm$^2$); $Tcc_{12}$ (ppm/°C) | 1.25; -180 |
| $c_{13}$ ($10^{11}$ dyn/cm$^2$); $Tcc_{13}$ (ppm/°C) | 1.2; -160 |
| $c_{33}$ ($10^{11}$ dyn/cm$^2$); $Tcc_{33}$ (ppm/°C) | 3.95; -100 |
| $c_{44}$ ($10^{11}$ dyn/cm$^2$); $Tcc_{44}$ (ppm/°C) | 1.18; -50 |
| $e_{31}$; $e_{33}$; $e_{24}$ (C/m$^2$) | -0.58; 1.55; -0.48 |
| $C_{66}$ | 1.1; 0 |
| $\varepsilon_{11}$; $\varepsilon_{33}$ ($10^{-11}$ F/m) | 9; 10.7 |
| ρ (kg/m$^3$); $\alpha_{11}$; $\alpha_{22}$; $\alpha_{33}$ (ppm/°C) | 3260; 5.27; 5.27; 4.15 |
| Si | |
| $c_{11}$ ($10^{11}$ dyn/cm$^2$); $Tcc_{11}$ (ppm/°C) | 1.66; -84 |
| $c_{12}$ ($10^{11}$ dyn/cm$^2$); $Tcc_{12}$ (ppm/°C) | 0.639; -93 |
| $c_{44}$ ($10^{11}$ dyn/cm$^2$); $Tcc_{44}$ (ppm/°C) | 0.796; -77 |
| $\varepsilon_{11}$ ($10^{-11}$ F/m) | 10.62 |
| ρ (kg/m$^3$); $\alpha_{11}$ (ppm/°C) | 2330; 0.87 |
| SiO$_2$ | |
| $c_{11}$ ($10^{11}$ dyn/cm$^2$); $Tcc_{11}$ (ppm/°C) | 0.785; 239 |
| $c_{12}$ ($10^{11}$ dyn/cm$^2$); $Tcc_{12}$ (ppm/°C) | 0.161; 548 |
| $\varepsilon_{11}$ ($10^{-11}$ F/m) | 3.32 |
| ρ (kg/m$^3$); $\alpha_{11}$ (ppm/°C) | 2200; 0.55 |

The ZGVRs admittance Y was investigated for different temperature values in order to calculate the TCF of each device. The real part (the conductance, G) and the imaginary part (the susceptance, B) of Y=G +jB were calculated at different temperatures in the -30 to 220°C range and some results are shown in figures 11a-f and 12a-f, as an example. Figures 11a-f show the conductance vs frequency curves of the ZGV3, ZGV5, ZGV7, ZGV9, ZGV10a and ZGV10b modes at temperature T = -30. 20, 70, 120, 170, and 220°C. Figures 12a-f show the susceptance vs frequency curves of the ZGV point of the ZGV3, ZGV5, ZGV7, ZGV9, ZGV10a and ZGV10b modes, for different temperatures (T = -30, 20, 70, 120, 170 and 220°C).

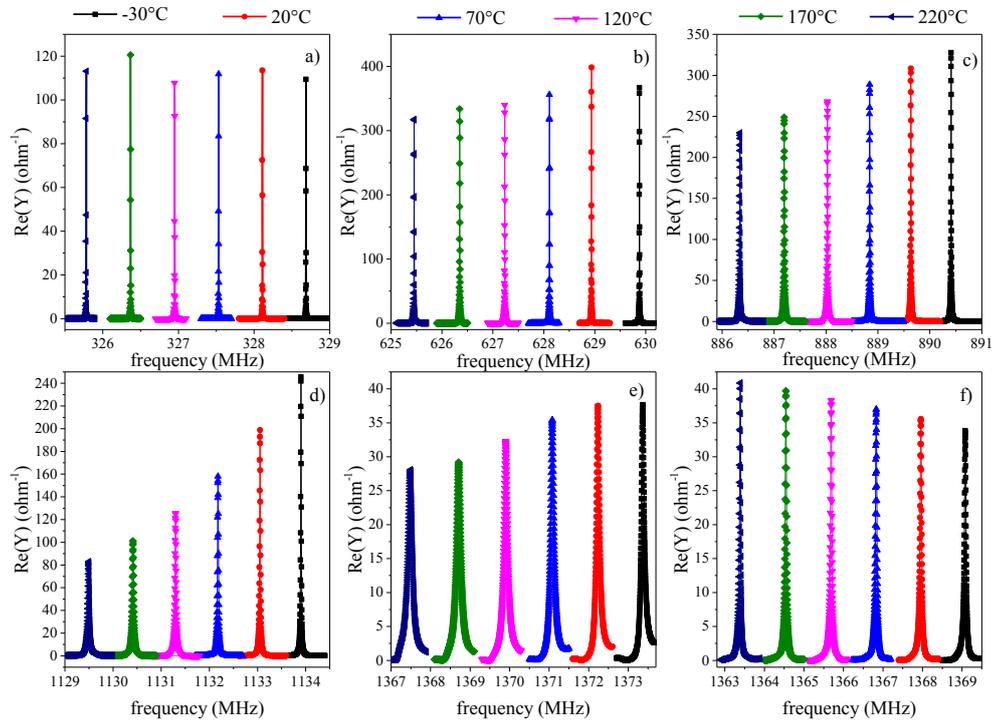

Figure 11: The real part of the admittance vs frequency curves of the a) ZGV3, b) ZGV5, c) ZGV7, d) ZGV9, e) ZGV10a, and f) ZGV10b modes at temperature T = -30. 20, 70, 120, 170, and 220°C.

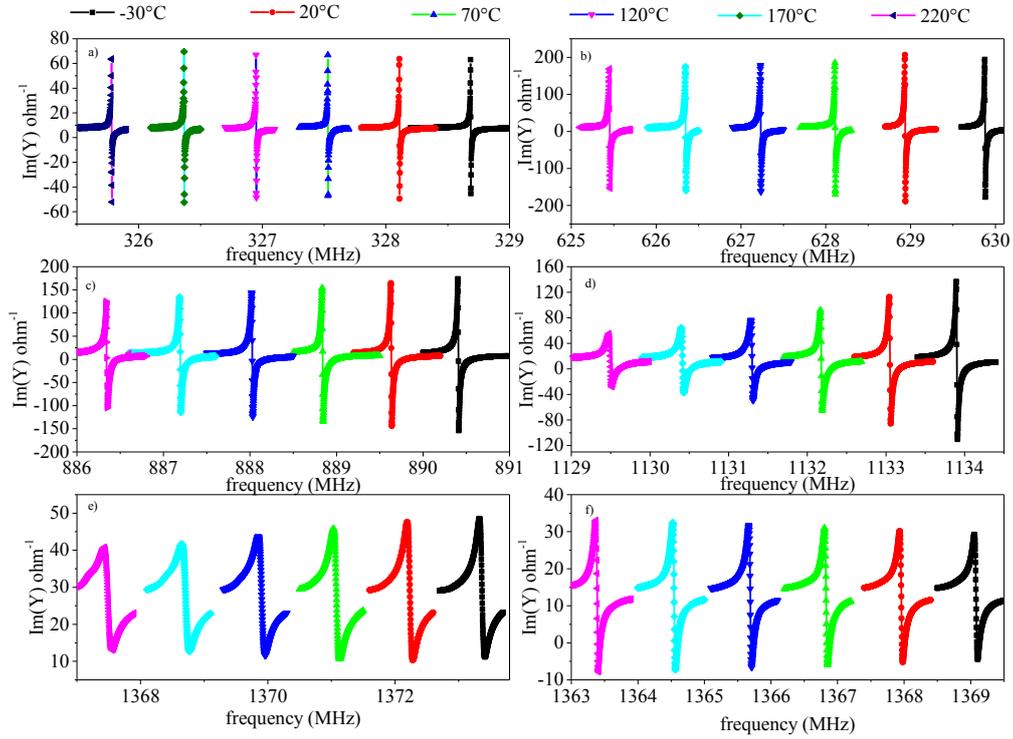

Figure 12: The imaginary part of the admittance Y vs frequency curves of the a) ZGV3, b) ZGV5, c) ZGV7, d) ZGV9, e) ZGV10a, and f) ZGV10b modes at temperature T = -30. 20, 70, 120, 170, and 220°C.

From figure 11a-f it can be seen that the frequency of the admittance of all the modes is shifted towards lower frequencies with increasing the temperature, thus confirming the negative sign of the TCF of all the ZGV points. The TCF values of the ZGV3, ZGV5, ZGV7, ZGV9, ZGV10a and ZGV10b modes resulted equal to -34, -27, -18, -16, -18 and -17 ppm/°C, respectively. The quality factor Q of the ZGVRs was estimated from the conductance vs frequency curves as the ratio between the resonant frequency $f_r$ and the bandwidth at half of the peak-conductance $\Delta f$, $Q = f/\Delta f$ [31]. The equivalent motional resistance $R_m$ was calculated as the inverse of the maximum admittance at resonance [31]. $R_m$ represents the acoustic energy loss of the resonator: in the ideal case it should be equal to zero if the displacement and the mechanical energy in the resonator were perfectly confined under the IDTs. The $R_m$ of the theoretically studied ZGVRs has a non null value even if very small, (approx.. 0.009, 0.0025, 0.0032, 0.005, 0.027, and 0.028 Ω) that is due to the choice of the wavelength value that doesn't perfectly match the value corresponding to zero group velocity. The wavelength values used in the present calculations were rounded to a value compatible with practical experimental device fabrication, as listed in table 1: hence a non-zero but extremely low group velocity (around 0.2 m/s) is to be expected together with a non-zero $R_m$ value. As a consequence, when the temperature changes, the wavelength mismatch worsens and the worsening depends on the mode order: the $R_m$ value changes thus indicating that the acoustic energy confinement is becoming slightly weaker even if still very low. At the same time the Q factor of the ZGVRs decreases with increasing the temperature.

## 5. Gas sensing application

At the ZGV points, the mode energy is locally trapped in the source area thus these modes are expected to be highly sensitive to the plate thickness and mechanical properties changes, as demonstrated in reference [32] for homogeneous isotropic plates. In the present study this method is applied to an inhomogeneous composite plate to investigate how the resonator characteristics are affected by the perturbation induced in the mass density of the outer plate layer. The behaviour of the ZGVRs operating as gas sensors was studied under the hypothesis that the surface of the device is covered with a thin polyisobutylene (PIB) film, 0.5 μm thick. The sensor was investigated for the detection of five volatile organic compounds at atmospheric pressure and room temperature: dichloromethane ($CH_2Cl_2$), trichloromethane ($CHCl_3$), carbontetrachloride ($CCl_4$), tetrachloroethylene ($C_2Cl_4$), and trichloroethylene ($C_2HCl_3$). The phase and group velocity dispersion curves of the PIB-covered SOI/AlN structure were calculated and it was found that, with respect to the uncoated SOI/AlN case, the presence of the PIB layer slightly lowers the frequencies of the first three ZGV points and induces new ZGV points, whose resonant frequencies in air, $f_0$, are listed in table 4. The interaction of the gas molecules with the sensitive PIB was simulated as an increase of the mass density of the PIB film, $\rho = \rho_{unp.} + \Delta\rho$, being $\rho_{unp.}$ the unperturbed mass density of the PIB layer (in air) and $\Delta\rho$ the partial density of the gas molecules adsorbed in the PIB layer, $\Delta\rho = K \cdot M \cdot c_0 \cdot P/RT$, where P and T are the ambient pressure and temperature (1 atm and 25°C), $c_0$ is the gas concentration in ppm, $K = 10^{1.4821}$ is the air/PIB partition coefficient for the studied gas, M is the molar mass, R is the gas constant [33-35]. Any effects of the gas adsorption on the PIB layer properties other than the density changes were neglected. The PIB gas adsorption was simulated for gas concentration $c_0$ in the range from 100 to 500 ppm for the ZGV3, ZGV5, ZGV7, ad ZGV8 modes. It was found that the ZGV resonance frequencies were downshifted by the adsorption of the gas into the PIB sensitive layer: the adsorbed gas increases the PIB mass density and lowers the phase velocity (and then the operating frequency), whch can be correlated to the gas concentration. Figures 10 a-d show the resonant frequency shift vs gas concentration for the four ZGV points; the corresponding resonant frequencies in air were equal to 323.1298, 616.667, 862.546, and 997.623 MHz, respectively. The resonant frequency shift of each sensor, $\Delta f = f_{air} - f_{c0}$, was calculated, being $f_{air}$ and $f_{c0}$ the resonat frequency values in air and at gas concentration $c_0$. As it can be seen, the frequency shift of each mode has a linear behaviour with respect to the increased mass density of the PIB layer. Moreover the slope of the curves (i.e., the sensor sensitivity) increases with increasing the resonant frequency.

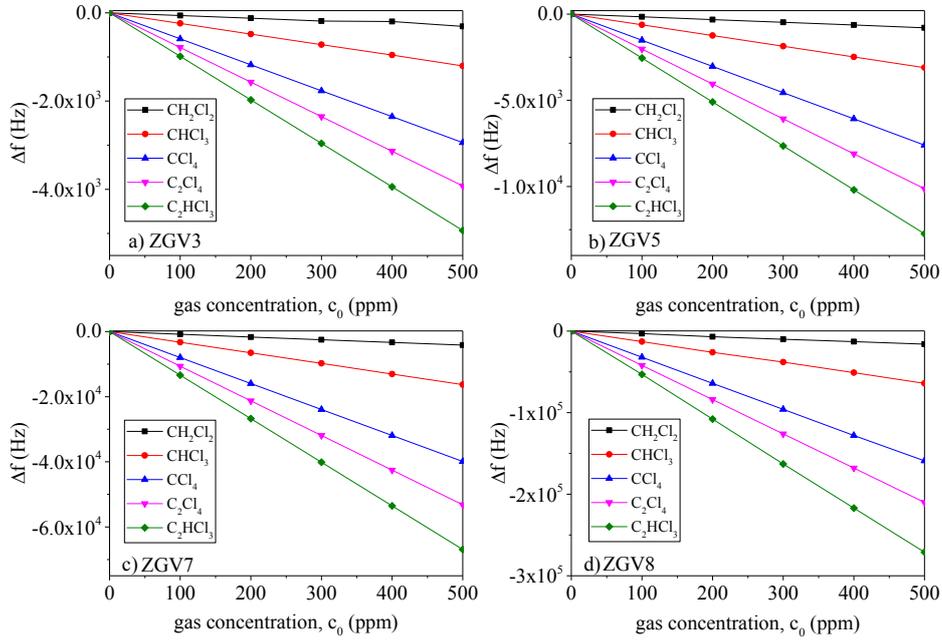

**Figure 13**: The resonant frequency shift vs gas concentration curves of the a) ZGV3, b)ZGV5, c) ZGV7, and d) ZGV8 mode for five different gases.

The ZGV sensors sensitivity to a fixed concentration ($c_0$ = 100 ppm) of the five volatile organic gases was compared with the theoretical sensitivity of a SAW sensor implemented on a yz-LiNbO$_3$ piezoelectric substrate covered by a PIB layer, 0.5 μm thick, with operating frequency equal to 1.121 GHz [36, 37]. Table 4 lists the operating frequencies in air, the PIB mass density increase $\Delta\rho$, and the frequency shifts of each ZGV mode anf of the SAW sensor of reference [36, 37] when exposed to 100 ppm of various gases at atmospheric pressure and temperature.

**Table 4**: The operating frequencies in air, $f_0$, the PIB mass density increase $\Delta\rho$, and the frequency shift $\Delta f$ of the ZGV modes in SOI/AlN and of the SAW in yz-LiNbO$_3$ when exposed to 100 ppm of various gases.

| | | $\Delta f$(Hz) | | | | |
|---|---|---|---|---|---|---|
| Gas (100 ppm) | PIB $\Delta\rho$ (kg/m$^3$) | ZGV3 $f_0$ = 323.1298 MHz | ZGV5 $f_0$ = 616.667 MHz | ZGV7 $f_0$ = 862.546 MHz | ZGV8 $f_0$ = 997.623 MHz | SAW in LiNbO3 $f_0$ = 1121.215 MHz |
| air | 0 | 0 | 0 | 0 | 0 | 0 |
| CH$_2$Cl$_2$ | 0.01 | -61 | -158 | -831 | -3000 | -356 |
| CHCl$_3$ | 0.04 | -240 | -621 | -3261 | -13000 | -1394 |
| CCl$_4$ | 0.1 | -588 | -1518 | -7975 | -32000 | -3408 |
| C$_2$Cl$_4$ | 0.132 | -784 | -2025 | -10635 | -42000 | -21831 |
| C$_2$HCl$_3$ | 0.166 | -986 | -2547 | -13378 | -53000 | -4544 |

To fully evaluate the potential of these sensors for detecting very low gas concentrations, temperature fluctuation effects should be removed from the sensor response. To this purpose some strategies can be exploited: 1. incorporation of a temperature sensor and compensation circuitry or software; 2. dual device configurations design; 3. temperature compensation. The first method is based on the numerical correction of the sensor response by using data from an independent measurement of the temperature by means of a nearby temperature sensor. The second method is based on the use of a dual resonator configuration that consists of a reference device and an active device: both the two devices are implemented on the same substrate and are covered by the PIB layer. The reference and active sensors must be configured to allow the former to be exposed both to the carrier gas and to the common-mode interfering measurands, while the latter is also exposed to the gas to be tested. A mixer circuit provides the difference between the signal from the reference device and that from the active device, so that the mass density variations of the PIB layer exposed to the gas could be distinguished independently of the effects of the temperature changes. Alternatively, a "stack" of metal and/or oxide layers of proper thicknesses and appropriate sign of the temperature coefficient of stiffness can be add to the device to yield a composite structure exhibiting minimal temperature coefficient [38].

## 6. Conclusions

The paper provides a theoretical analysis on the several ZGV modes for a layered medium comprising an AlN piezoelectric layer and two non-piezoelectric layers of $SiO_2$ and Si. The Lamb-like modes propagation along a supported thin three-layers structure was studied and the frequencies corresponding to zero group velocity were found. The mode shape, the coupling efficiency and the phase velocity of the ZGV points were theoretically estimated. The thermal behaviour of the acoustic waveguide was studied in the -30 to 220°C range and a negative.

The present paper also theoretically investigates the applicability of the ZGVRs for gas sensing application. The resonators surface was covered with a sensitive PIB layer that selectively adsorpt the gas molecules. As a result of the modified surface loading, the device changes its operation frequency. The sensitivity to five different gases with concentration in the 100 to 500 ppm were calculated for four ZGV points and compared with a SAW sensor implemented on LiNbO3 and covered with a same thickness PIB membrane. The high Q and low loss of the ZGVR-based sensors are remarkable features that can result in substantial noise reduction and resolution improvement. The feasibility of a one-port ZGV Lamb wave resonator on SOI/AlN substrate, that is realistic to fabricate and achieves a range of phase velocity spanning from about 16000 to 47000 m/s, with a $K^2$ in the range from 0.03 to 0.09% was demonstrated theoretically.

## 7. Acknowledgment

This project has received funding from the European Union's Horizon 2020 research and innovation programme under the Marie Sklodowska-Curie Grant Agreement No. 642688.